\documentclass[11pt,a4paper]{article}
\pdfoutput=1
\usepackage{jheppub}

\usepackage{graphicx}
\graphicspath{ {images/} }
\usepackage{amsmath,amssymb,amsthm}
\usepackage{latexsym,graphicx,color,subfigure}
\usepackage{bm}

\newcommand{\be}{\begin{equation}}
\newcommand{\ee}{\end{equation}} 
\newcommand{\beq}{\begin{eqnarray}}
\newcommand{\eeq}{\end{eqnarray}}

\newcommand{\D}{\mathcal{D}}
\newcommand{\p}{\partial}
\newcommand{\Tr}{{\rm Tr}}

\newcommand{\bea}{\begin{eqnarray}}
\newcommand{\eea}{\end{eqnarray}}
\def\Tr{ \hbox{\rm Tr}}

\def\half{\frac{1}{2}}

\def\half{\frac{1}{2}}

\title{{\bf BPS Alice strings }}
\author{Chandrasekhar Chatterjee$^{1, a}$} 
\author{Muneto Nitta$^{1,b}$}
 \affiliation[a]{Department of Physics, and Research and Education Center for Natural Sciences,\\ Keio University,Hiyoshi 4-1-1, Yokohama, Kanagawa 223-8521, Japan}
 \emailAdd{chandra.chttrj@gmail.com$^{a}$, chandra@phys-h.keio.ac.jp$^{a}$}
 \emailAdd{nitta(at)phys-h.keio.ac.jp$^{b}$}

\date{\today}

\abstract{
When a charged particle encircles around an Alice string, it changes the sign of the electric charge.  In this paper we find a BPS-saturated Alice string in 
 $U(1)\times SO(3)$ gauge theory with charged complex scalar fields 
 belonging to the vector representation. 
After performing BPS completion we solve the BPS equations numerically. 
We further embed the Alice string into 
an ${\cal N}=1$ supersymmetric gauge theory 
to show that it is half BPS.
}

\begin{document}
\maketitle
\flushbottom

\section{Introduction}
An Alice string was discovered theoretically almost four decades ago by Albert S. Schwarz \cite{Schwarz:1982ec}.  
This string has a non-trivial topological property that 
charged particles flip the sign of the electric charge 
after encircling the Alice string once,
and consequently the charge is not globally defined.  
The theory admitting an Alice string is sometime called as 
``Alice electrodynamics", which is  a  $U(1)$ gauge  theory that includes the charge conjugation as a local symmetry,  
as was discussed first time 
even before by Joe Kiskis \cite{Kiskis:1978ed}.
A typical Alice string was found in an $SO(3)$ gauge theory 
with scalar fields in spin-2 (traceless symmetric tensor) representation of $SO(3)$.
The $SO(3)$ gauge group is spontaneously broken to $O(2)$, 
giving rise to the vacuum manifold or order parameter space 
$G/H =SO(3)/O(2) \simeq {\mathbb R}P^2$.
The non-trivial homotopy group 
$\pi_1({\mathbb R}P^2) \simeq {\mathbb Z}_2$ admits an Alice string.
Then, the unbroken generator, which we identify as the electromagnetic $U(1)$ generator,  
changes its sign when encircling the Alice string once.  A possible  application of Alice strings in cosmology was proposed in ref.~\cite{Blinnikov:1982eh}.

   The Alice string is an example of a wide class of systems of spontaneously broken symmetry where the unbroken symmetry is either  a discrete group or a continuous group that contains components which are not connected to the identity.   In the case of the Alice string, the unbroken gauge group  is $O(2)$ which is a continuous group with more than one connected components. 
    In general, embedding  of the unbroken group becomes  space  dependent inside the full symmetry group in the presence of vortices. The presence of discrete symmetry or disconnected components in the unbroken gauge group often makes some of the generators undefined globally \cite{Alford:1990mk, Alford:1990ur, Alford:1992yx}. In the case of the Alice string, the generator of unbroken $U(1)$ group changes 
    the sign after encircling once around the vortex.   This shows a very interesting exotic phenomenon  called the creation of ``Cheshire charge'', a charge which is lost after charged particles  encircle the string once \cite{Preskill:1990bm, Bucher:1993jj, Bucher:1992bd}.  This allows  all kinds of interesting global effects, such as the generalized Aharonov-Bohm effect, anyonic exchange statistics  etc \cite{Lo:1993hp}. Not only electric Cheshire charge, the Alice string also exchanges magnetic charges from magnetic particles, creating the Cheshire magnetic charges \cite{Bucher:1992bd}. Interesting thing is that a ring of the Alice string can be turned into a magnetic monopole at large distances once the $U(1)$ modulus is twisted along the ring \cite{Shankar:1976un,  Benson:2004ue, Bais:2002ae, Striet:2003na}.      
 This follows from the fact that  in case of the Alice string, the spontaneous symmetry breaking gives non-trivial second homotopy group, $\pi_2(G/H)\ne 0$ .    
    
    On the other hand, Bogomol'nyi-Prasad-Sommerfield (BPS) 
    states are the most stable states among configurations with 
    a fixed topological charge 
    since they saturate Bogomol'nyi bound of the energy 
    \cite{Bogomolny:1975de, Prasad:1975kr}. 
    In this case, the total energy is proportional to the topological 
    charge, namely the number of solitons.
    Abrikosov-Nielsen-Olesen (ANO) vortices (flux tubes) 
    \cite{Abrikosov:1956sx,Nielsen:1973cs}  
    become BPS at the critical coupling between type-I and type-II superconductors. 
    Such BPS vortices preserve a half supersymmetry (SUSY)
    when embedded into SUSY gauge theories. 
    They belong to so-called short SUSY multiplets and 
    consequently are stable against quantum corrections \cite{Witten:1978mh}. 
    Therefore they play crucial roles to determine 
    non-perturbative effects of the theories.
   BPS non-Abelian vortices are also present in certain 
SUSY non-Abelian gauge theories  
\cite{Hanany:2003hp, Auzzi:2003fs, Shifman:2004dr, 
Hanany:2004ea, Gorsky:2004ad, 
Eto:2005yh, Eto:2006cx, Eto:2006db, Eto:2006pg}.  
On the other hand,   the symmetry breaking which creates conventional 
    Alice strings \cite{Schwarz:1982ec} gives the fundamental group $\pi_1(G/H) \simeq \mathbb{Z}_2$,  
    implying that an Alice string can annihilate itself when two of them collide, contradict to a BPS property that energy 
    is proportional to the topological charge.
     Therefore, Alice strings are naturally non-BPS.\footnote{A BPS Alice string in ${\mathcal N}=4$ SUSY gauge theory was discussed without an explicit construction in string theory \cite{Harvey:2007ab, Harvey:2008zz}.
     There is also an attempt in Ref~\cite{Striet:2000bf} to obtain a BPS Alice string of the conventional type but no Bogomol'nyi completion could be achieved, 
    where the system is forced to  become of first order by assuming  some constraints  by hand in a particular gauge.}
    
   Another type of Alice strings is known in Bose-Einstein condensates (BEC) \cite{Leonhardt:2000km}, see also 
   \cite{Kobayashi:2011xb,Kawaguchi:2012ii}.
 More presicely, the theory is a spinor BEC with scalar fields belonging 
 to vector (spin-1) representation of $SO(3)$, 
 for which a group $U(1) \times SO(3)$ is spontaneously broken to 
 $O(2)$ giving an order parameter space 
 $G/H = [U(1) \times SO(3)] /O(2) \simeq 
 [S^1 \times S^2] /\mathbb{Z}_2$ with 
 a non-trivial homotopy group 
 $\pi_1(G/H) \simeq \mathbb{Z}$. 
 This admits a global analog of Alice strings but one does not 
 have a charged particle since all symmetries $G$ are global 
 symmetries.
 
 In this paper, we find that a BPS Alice string exists 
 in a $U(1) \times SO(3)$ gauge theory 
 coupled with charged complex scalar fields in 
 the vector representation.
 This is achieved simply by gauging the symmetry $G=U(1) \times SO(3)$ 
 in the corresponding spin-1 BEC model \cite{Leonhardt:2000km}.
We numerically construct an axially symmetric Alice string configuration,
and find that equations for profile functions for it coincide 
with those for a non-Abelian vortex \cite{Auzzi:2003fs} 
in $U(N)$ gauge theory coupled with $N$ complex Higgs fields 
in the fundamental representation.
We further embed our bosonic model into an ${\cal N}=1$ SUSY gauge theory 
and show that the BPS Alice string is a 1/2 BPS state,
preserving a half SUSY.

 This paper is organized as follows.    In Sec.~\ref{sec:2} 
 we introduce a simple model of $U(1)\times SO(3)$ gauge theory with a complex scalar field in a vector representation. 
 We perform the Bogomol'nyi completion of the energy functional and write down the BPS equations. In Sec.~\ref{sec:sol}  
we introduce an ansatz and boundary condition for a single Alice string. 
We expand BPS equations in the profile functions and solve the equations numerically. 
In Sec.~\ref{sec:sym} we discuss 
a symmetry structure of our Alice string. 
Section \ref{sec:susy} is devoted to a summary 
and discussion.

\section{The model for BPS Alice strings
and the BPS equations}\label{sec:2}
We consider an $SO(3) \times U(1)$ gauge theory 
coupled with charged complex scalar fields in the vector representation,
equivalently an $ SU(2)\times U(1)$ gauge theory 
coupled with 
one charged  complex scalar field 
in the adjoint representation. Here we use the latter expression for usefulness. 
 We denote $SU(2)$ and  $U(1)$ gauge fields by
$A_{\mu}=A_{\mu}^a \tau^a$ 
and $a_{\mu}$, respectively, 
and a charged  complex scalar field 
in the 
adjoint representation by  
 $\Phi = \Phi^a \tau^a$, where 
 $\tau^a = \half \sigma^a$ 
 and $\sigma^a(a = 1,2,3)$ are the Pauli matrices. 
The action is given by
 \begin{eqnarray}
 \label{action}
&&{I} = \int d^4x \left[- \frac{1}{2} \Tr F_{\mu\nu} F^{\mu\nu}- \frac{1}{4} f_{\mu\nu}f^{\mu\nu} + \Tr | D_\mu\Phi|^2  
-  \frac{\lambda_g}{4} \Tr[\Phi,\Phi^\dagger]^2 
-  \frac{\lambda_e}{2}\left(\Tr \Phi\Phi^\dagger  - 2 \xi^2\right)^2\right]. \nonumber \\ \end{eqnarray}
The covariant derivatives and field strengths are defined by
$D_\mu\Phi = \p_\mu\Phi - i  e a_\mu\Phi -i g \left[A_\mu, \Phi\right]$, 
and 
$F_{\mu\nu} = \p_\mu A_\nu - \p_\nu A_\mu - i g [A_\mu, A_\nu], \, f_{\mu\nu} =  \p_\mu a_\nu - \p_\nu a_\mu$,  
with the gauge couplings $g$ and $e$ for 
$SU(2)$ and  $U(1)$ gauge fields, respectively. 
Since the center of $SU(2)$ does not acts on the adjoint field,
the gauge symmetry is actually 
$G = U(1)_b \times \frac{SU(2)}{\mathbb{Z}_2} \simeq U(1)_b \times SO(3)$.

In the vacuum, the scalar field takes the non-zero vacuum expectation value as
\begin{eqnarray}
\langle\Phi\rangle = 2 \xi \tau^1,\;
\end{eqnarray}
which keeps the $U(1)_1$ symmetry generated by $\tau^1$ 
unbroken and consequently $A_\mu^1$ remains massless, 
while other gauge fields become massive. 
 The breaking pattern of the gauge symmetry is
 \begin{eqnarray}\label{symmetry}
G = U(1)_b \times \frac{SU(2)}{\mathbb{Z}_2} \simeq U(1)_b \times SO(3) \longrightarrow  H \simeq 
 \mathbb{Z}_2 \ltimes  U(1)_1 \simeq O(2)\;
\end{eqnarray}
where $\ltimes$ denotes a semi-direct product. 
The unbroken $\mathbb{Z}_2$ is given by 
a simultaneous action of 
a $\pi$ rotation around either $\tau^3$, 
$\tau^2$ or their linear combination 
and a $\pi$ rotation in $U(1)_b$.
Since the $\mathbb{Z}_2$ action does not commute with 
$U(1)_1$, there is the semi-direct product $\ltimes$ between them.
The vacuum manifold is obtained as
\begin{eqnarray}
 {G\over H} 
 = \frac{U(1)_b\times SO(3)}{O(2)}
 \simeq \frac{S^1 \times S^2}{\mathbb{Z}_2}.\label{eq:vmfd}
\end{eqnarray}
This allows a non-trivial fundamental group 
\begin{eqnarray}
\pi_1  \left( G/H\right)
\simeq \mathbb{Z},
\end{eqnarray}
indicating the existence of stable vortices. 
One can observe that compared with 
the vacuum manifold $S^2/{\mathbb{Z}_2} \simeq {\mathbb{R}}P^2$ 
of the conventional Alice 
string, there is the $U(1)$ factor in Eq.~(\ref{eq:vmfd}).
We will see that this difference is essential for the BPS property.

   The static energy is expressed as 
\begin{eqnarray}
\mathcal{H} &=& \int d^3x \left[\frac{1}{2} \Tr F_{ij}^2 + \frac{1}{4} f_{ij}^2 + \Tr | D_i\Phi|^2 +   \frac{\lambda_g}{4} \Tr[\Phi,\Phi^\dagger]^2
 +  \frac{\lambda_e}{2}\left(\Tr \Phi\Phi^\dagger - 2 \xi^2\right)^2 \right].
\end{eqnarray}
Here, we consider the critical couplings,
\begin{eqnarray}
 \lambda_e = e^2, \quad \lambda_g = g^2
\end{eqnarray} 
at which we can consider the Bogomol'nyi's arguments.
By performing the Bogomol'nyi completion, 
the tension (energy per the unit length) of a vortex 
along the $x_3$ coordinate
can be written as
\begin{eqnarray}
\mathcal{T} & =& \int d^2x \left[\Tr\left[ F_{12} \pm \frac{g}{2}[\Phi,\Phi^\dagger]\right]^2 + \Tr|\D_{\pm} \Phi|^2 + \half\left[  f_{12} \pm e \left(\Tr \Phi\Phi^\dagger - 2 \xi^2\right)\right]^2
\pm 2 e f_{12} \xi^2 \right.\Big{]} \nonumber\\ 
 &\ge& 2 e  \xi^2 \left|\int d^2x\; f_{12}\right|,\;
\end{eqnarray}
where $\D_{\pm} \equiv \frac{D_1 \pm i D_2}{2}$ are  used together with 
the complex coordinate $z \equiv x_1 + i  x_2$, 
and all fields are taken to be independent of $x_3$ coordinate.
The BPS condition is satisfied when the inequality is saturated, 
in which case the tension is simply given by
\begin{eqnarray}
\mathcal{T}_{\rm BPS} =  2 e  \xi^2 \left| \int d^2x f_{12}\right|.
\end{eqnarray}
At this saturation point we can write first order BPS equations 
\begin{eqnarray}
&&
 f_{12} \pm e \left(\Tr \Phi\Phi^\dagger - 2 \xi^2\right) =0,
 \label{eq:BPS1}
 \;\\
&&F_{12} \pm \frac{g}{2}[\Phi,\Phi^\dagger] =0,\;
\label{eq:BPS2}
\\
&& \D_{\pm} \Phi  = (\D_{\pm} \Phi)^\dagger   = 0.\label{eq:BPS3}
\end{eqnarray}
All configurations that satisfy the above BPS equation 
also satisfy the full second order equations of motion as usual.

\section{A BPS Alice string solution}\label{sec:sol}
\begin{figure}[!htb]
\centering
\subfigure[\, ]{\includegraphics[totalheight=3.5cm]{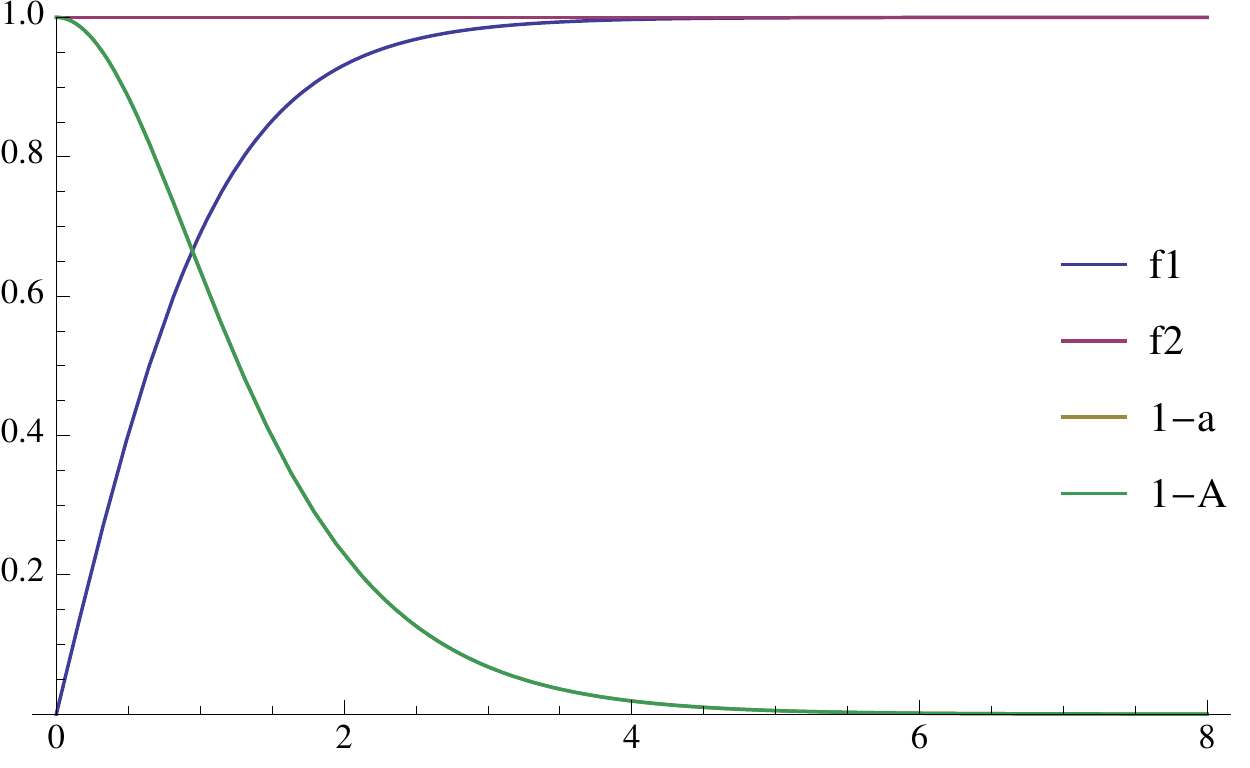}}
\subfigure[\, ]{\includegraphics[totalheight=3.5cm]{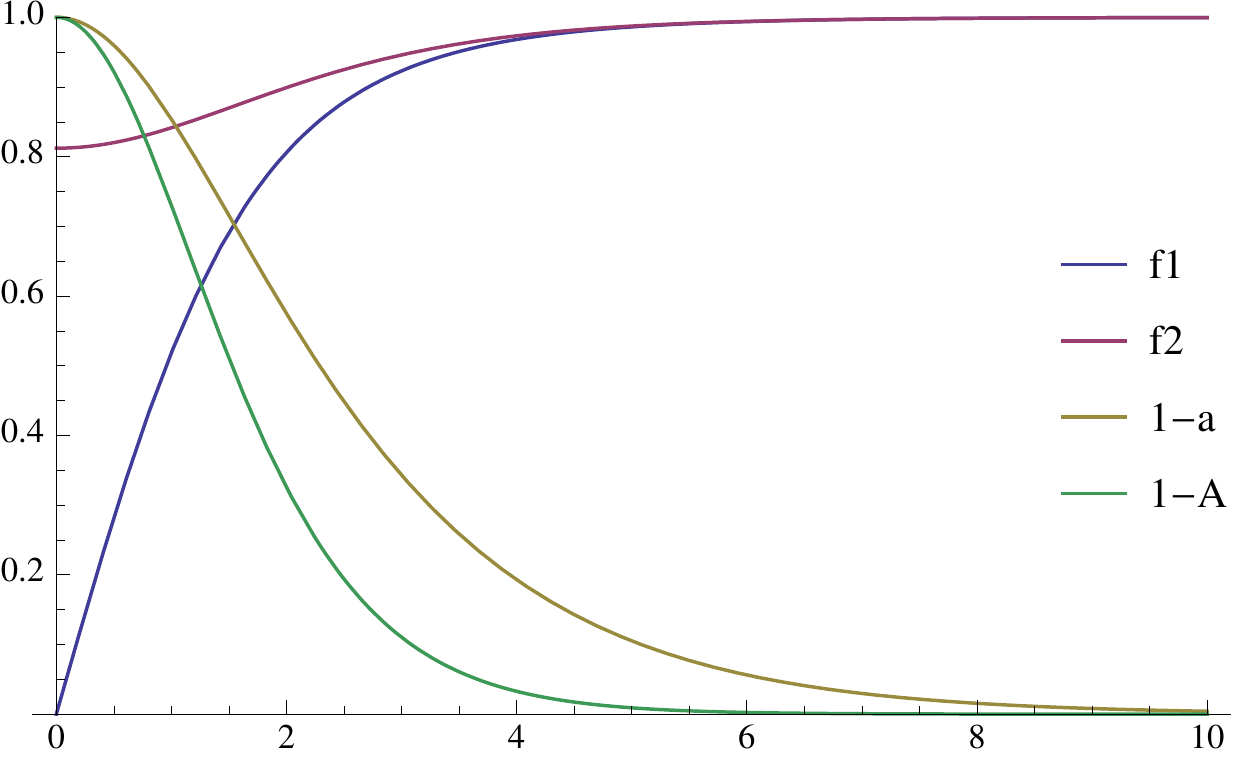}}\\
\subfigure[\, ]{\includegraphics[totalheight=3.5cm]{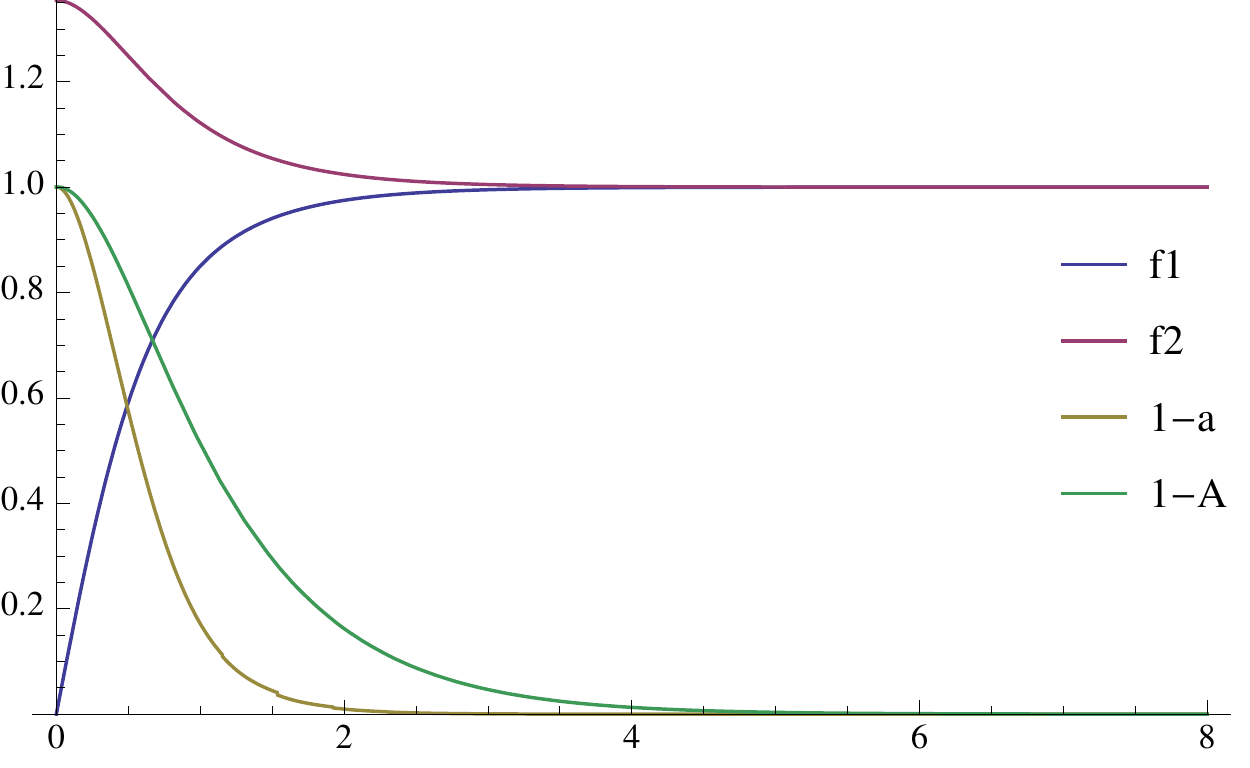}}
\caption{\em\small The profile functions $f_1(\rho)$ ,  $f_2(\rho)$, $1 -a(\rho)$ and $1-A(\rho)$ 
as functions of the distance  $\rho$  from the vortex center 
 for a single Alice string vortex. 
 We plot them for different values of the ratio $l = \frac{e}{g}$ 
 between the $U(1)$ and $SO(3)$ gauge couplings.  
 (a) $l$ =1, (b) $ l = \half $, (c) $ l = 5 $. }
\label{profile}
\end{figure}

In this section, we construct a cylindrically symmetric single 
vortex solution.
To solve the BPS equations, 
we consider the ansatz 
\begin{eqnarray}
&&\Phi(r, \varphi) = 
\xi\left(
\begin{array}{ccc}
 0 & f_1(r) e^{i\varphi}     \\
f_2(r)   & 0  \end{array}
\right),\\
&&a_i (r, \varphi) = - \frac{1}{2e} \frac{\epsilon_{ijx_j}}{r^2}a(r), \;
A_i(r, \varphi) = -\frac{1}{4g} \frac{\epsilon_{ijx_j}}{r^2}\sigma^3A(r).
\end{eqnarray}
where $\{r, \varphi \}$ are radius and azimuthal angle, 
respectively. 
Here, 
$f_1(r), f_2(r), A(r)$ and $a(r)$ are four profile functions depending 
only on the radial coordinate.
Then, the first order equations take the form
\begin{eqnarray}
&& \frac{1}{r}\p_r a(r) + 2e^2 \xi^2\left[f_1(r)^2 + f_2(r)^2 - 2\right] =0,
\label{eq:profile1}\\ 
&& \frac{1}{r}\p_r A(r) + 2g^2 \xi^2\left[f_1(r)^2 - f_2(r)^2 \right] =0,\;\\
&& \p_r f_1(r) - \frac1r\left(1 - \frac{a(r) + A(r)}{2}\right) f_1(r) = 0 ,\\   
&& \p_r f_2(r) + \frac{a(r) - A(r)}{2 r} f_2(r) = 0 .\label{eq:profile4}
\end{eqnarray}
We solve the above equations 
with the boundary conditions
\begin{eqnarray}
f_1(0) = f_2'(0) = 0,&& f_1(\infty) = f_2(\infty) =1,\\ A(0)= a(0) = 0,&& A(\infty)= a(\infty) = 1.
\end{eqnarray}
To solve these equations let us define two new profile functions $\psi_0(r)$ and $\psi_1(r)$ as
\begin{eqnarray}
 f_1(r) = r e^{-\half(\psi_0(r)+ \psi_1(r))}, \; 
 f_2(r) =  e^{\half( \psi_1(r)-\psi_0(r))}.
\end{eqnarray}
These give
\begin{eqnarray}
A(r) =  r\psi'_1(r), \;  a(r) = r \psi'_0(r).
\end{eqnarray}
Then, 
the above four equations are reduced to the following two 
equations, 
\begin{eqnarray}
&& \frac{1}{\rho}\p_{\rho}\left[ \rho\psi_0' (\rho)\right]+ l^2\left[e^{-\psi_0}\left(\rho^2e^{-\psi_1}  + e^{\psi_1}\right) - 2\right] = 0,\\
&& \frac{1}{\rho}\p_{\rho}\left[ \rho\psi_1'(\rho)\right] + e^{-\psi_0}(e^{-\psi_1}  -  e^{\psi_1} ) =0,\;
\end{eqnarray}
where we have defined $ \rho^2  \equiv 2g^2 \xi^2 r^2$ and  
the ratio $l \equiv e/g$ between the $U(1)$ and $SO(3)$ gauge couplings. 
The boundary conditions for the new variables become
\begin{eqnarray}
\psi'_0(0) = \psi'_1(0),\; 
\psi_0(R) = \psi_1(R) =\log R ,\;
\end{eqnarray}
where $R$ is the system size. 
We solved these equations numerically and all the profile functions are plotted in Fig.~\ref{profile}.

We may  further rewrite the equations by defining 
\begin{eqnarray}
\tilde\psi_1 = \psi_1 - \log \rho\, \quad\tilde\psi_0 = \psi_0 - \log \rho 
\end{eqnarray}
and these redefinitions transform the equations to the form
\begin{eqnarray}
\frac{1}{\rho}\p_{\rho}\left[ \rho\tilde\psi_0'(\rho) \right]+ 2 l^2\left[e^{-\tilde\psi_0}\cosh \tilde\psi_1 - 1\right] &=& - 2\pi \delta^2(\rho),\;\\
 \frac{1}{\rho}\p_{\rho} \left[\rho\tilde\psi_1'(\rho) \right] - 2  e^{-\tilde\psi_0} \sinh \tilde\psi_1 &=& - 2 \pi\delta^2(\rho).
\end{eqnarray}
For $l = 1$ these two equations posses  the solution $\psi_1(\rho) = \psi_0(\rho)$, 
and the above two equations are reduced to 
a single equation, which is known as 
the Taubes equation \cite{Taubes:1979tm} for a single ANO vortex \cite{Nielsen:1973cs}.

Before closing this section, 
it may be interesting to note that 
Eqs.~(\ref{eq:profile1})--(\ref{eq:profile4}) for profile functions
are identical to those 
of an axially symmetric non-Abelian vortex 
in $U(N)$ gauge theory coupled with 
$N$ flavors in the fundamental representations 
\cite{Auzzi:2003fs}.
Although the physical properties are quite different between them,
some mathematical properties such as the uniqueness and existence of the solutions are common.

\section{Symmetry structure of the Alice string}\label{sec:sym}
 In this section we shall analyze the symmetries and symmetry breaking  of the system.
To understand the symmetry breaking defined in Eq.~(\ref{symmetry}), let us  analyze the large distance behavior of  the order parameter 
of the scalar field. The order parameter has a non-trivial winding at large distances of the string and can be expressed as 
\begin{align}
\Phi(R, \varphi) & \sim \xi e^{i\frac{\varphi}{2}} \left(
\begin{array}{ccc}
 0 &  e^{i\frac{\varphi}{2}}     \\
  e^{-i\frac{\varphi}{2}} & 0  \end{array}
\right) 
= \xi e^{i\frac{\varphi}{2}} e^{i\frac{\varphi}{4}\sigma^3} \sigma^1 e^{-i\frac{\varphi}{4}\sigma^3}
\end{align}
with the system size $R$. 
Now if we  set the order parameter at $\varphi =0$ (along the $ x_1$-axis) 
as
\begin{eqnarray}
\Phi(R, \varphi=0) = \xi \sigma^1,\;
\end{eqnarray}
 the order parameter at any arbitrary $\varphi$
 can be obtained by a holonomy action as 
\begin{align}
\Phi(R, \varphi) &= e^{ie \int {\bf a\cdot dl}} e^{i g \int {\bf A\cdot dl}}\, \Phi(R, 0)e^{-ig \int {\bf A\cdot dl}} \nonumber\\
 & = U_0(\varphi) U_3(\varphi)\, \Phi(\infty, 0) U^{-1}_3(\varphi),\;
\end{align}
where we have defined holonomies by 
\begin{eqnarray}
\label{holonomy}
U_0(\varphi) = e^{ie \int_0^\varphi {\bf a\cdot dl}} = e^{i\frac{\varphi}{2}}, \quad U_3(\varphi) = Pe^{i g \int_0^\varphi {\bf A\cdot dl}} = e^{i\frac{\varphi}{4}\sigma^3} .
\end{eqnarray}
These are obtained by the condition that 
the order parameter $\Phi$ is covariantly constant at large distances ($\D_i\Phi \rightarrow 0$ as $R \rightarrow \infty$).
\begin{figure}[h]
\includegraphics[width= 8cm]{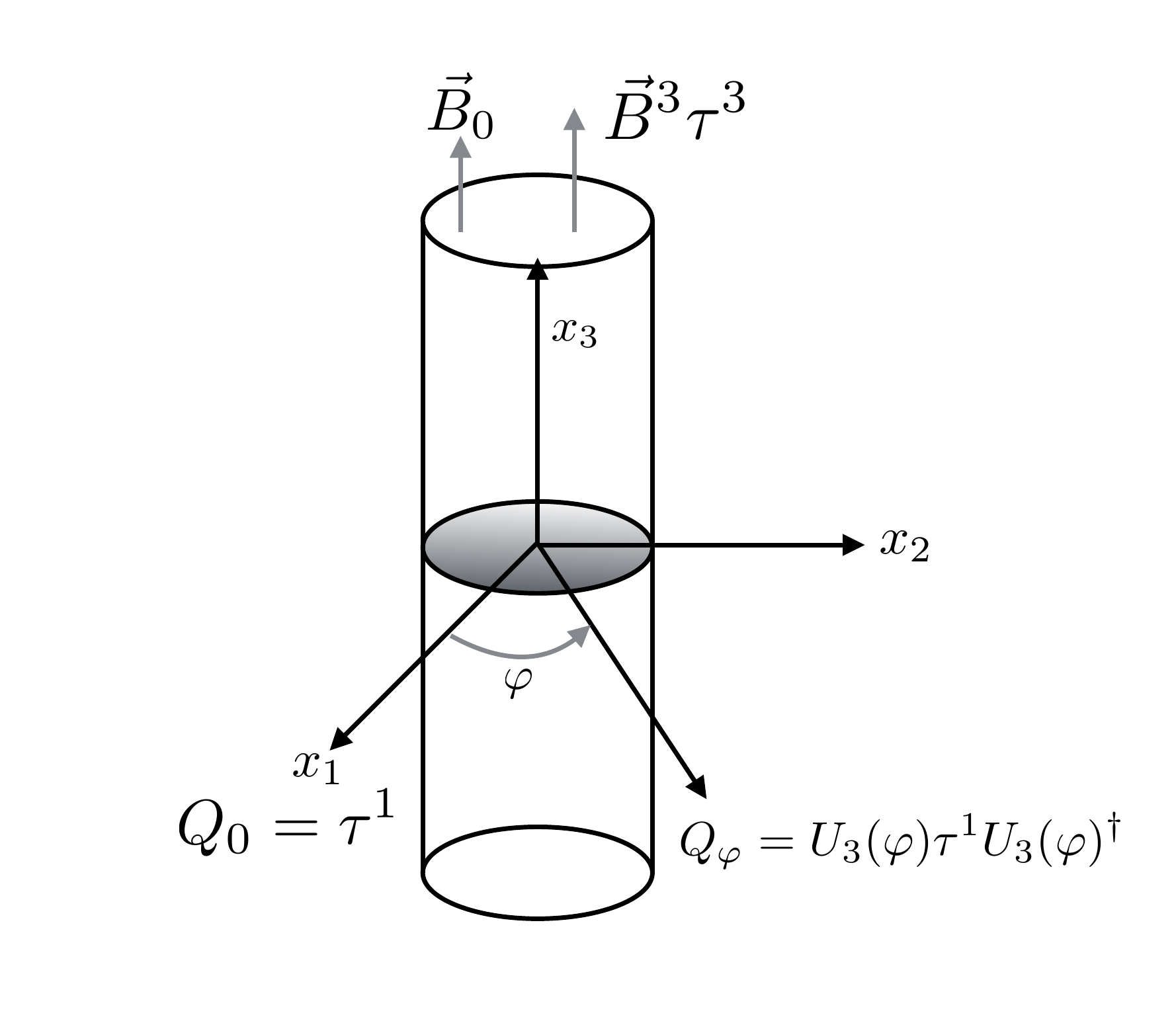}
\centering
\caption{\em\small The Alice string is schematically shown. The Abelian and non-Abelian magnetic fields along the $x_3$ axis  are denoted by $\vec{B}^0$ and $\vec{B}^3$, respectively.  The change of the electric charge of the unbroken $U(1)$ is  displayed by denoting $Q_0$ at $\varphi = 0$ and $Q_\varphi$ for arbitrary azimuthal angle outside the  string.}
\label{Alice string}
\end{figure}

Now it can be easily noticed that the generator of unbroken group changes as we encircle the string by an angle $\varphi$. 
The generator of the unbroken $U(1)$ at $\varphi=0$ 
is defined as $Q_0 = \tau^1$.  
Then the $U(1)$ generator at the angle $\varphi$ can be obtained 
by an action of the holonomy $U_3(\varphi)$ on $Q_0$ as
\begin{eqnarray}
Q_\varphi = U_3(\varphi) Q_0 U_3(\varphi)^{-1}, \quad U_3(\varphi) = e^{\frac{i\varphi}{4}\sigma^3}.
\end{eqnarray}
We then find that the generator changes its sign after the completion of one full 
circle around the vortex as
\begin{eqnarray}
Q_{2\pi} = - Q_0.
\end{eqnarray}
This makes the non-existence  of electric charge globally, 
as the case of conventional Alice strings.

The action of holonomy in Eq.~(\ref{holonomy}) 
can be written in the
$SO(3)$ spin one representation as
\begin{eqnarray}
J_3 = 
\left(
\begin{array}{ccc}
 \,1 & \,0  & \, 0 \\
\, 0 &\, 0   &\, 0  \\
\,0  & \,0  & -1  
\end{array}
\right),\; U_3(\varphi) = e^{\frac{i\varphi}{2}J_3}, \; U_3(0) = \bm{1}, \; U_3(2\pi) =  {\rm diag}.(-1, 1, -1).
\end{eqnarray}
This shows that
\begin{eqnarray}
U_3(2\pi) \in \mathbb{Z}_2 \subset O(2),
\end{eqnarray}
implying the Alice property more clearly.

Before closing this section, let us make a remark. The above Alice phenomenon is an example of ``obstruction"\cite{Nelson:1983bu, Abouelsaood:1982dz, Balachandran:1982gt, Bolognesi:2015ida}. In general,  a covariantly constant embedding of the unbroken symmetry group $H$ (the little group of $\Phi(\varphi)$)  inside the original symmetry group $G$ depends on $\varphi$. 
If $H$  is a non-Abelian discrete group or a continuous group containing discrete elements as a semi-direct product, some generators become multivalued. Such a system in which a gauge group contains disconnected elements was discussed in Ref.~\cite{Kiskis:1978ed} in the context of $O(2)$ gauge  theory  where charge conjugation symmetry($\mathbb{Z}_2$) was introduced as a local symmetry along with the $SO(2)$ gauge theory in 
a non-simply connected space.

\section{The  SUSY model and 1/2 BPS Alice strings}\label{sec:susy}

In this section, we embed the bosonic action in the last sections into
an ${\cal N} = 1$ SUSY gauge theory in $3 + 1$ dimensions 
(or equivalently ${\cal N}=2$ SUSY theory 
in $ 2 +1$ dimensions). Then, we show that BPS equations 
can be obtained by a half SUSY preserving condition.

 Let us introduce chiral superfields $\Psi_\pm (x,\theta,\bar\theta)$ 
transforming under the adjoint representation of the $SU(2)$ group 
with the $U(1)$ charge $\pm 1$,
and  the vector superfields  $V_0(x,\theta,\bar\theta)$ and $V(x,\theta,\bar\theta)=V^a (x,\theta,\bar\theta) \tau^a$ for 
the $U(1)$ and $SU(2)$ gauge groups, respectively. 
Then, 
the ${\cal N} = 1$ SUSY action that we consider is given 
in terms of the above superfields by
\begin{eqnarray}
&& \mathcal{I} = 
- \frac{1}{4}\Re\int d^4x d^2\theta
\left( 2 \Tr W^\alpha W_\alpha + W_0^\alpha W^0_\alpha\right) \nonumber\\
&& \hspace{0.8cm}
+\frac{1}{4} \int d^4x d^2\theta d^2\bar\theta 
\Tr \left(\Psi^\dagger_+ e^{-2eV_0 - 2g V} \Psi_+ 
+ \Psi^\dagger_- e^{2eV_0 - 2g V} \Psi_- 
- \xi^2 V_0
\right)
\end{eqnarray}
where $\xi^2$ is a constant called 
the Fayet-Iliopoulos parameter\footnote{The cosmic strings in the same theory were also discussed before in ref.~\cite{Davis:1997ny}. However explicit Bogomol'nyi completion and BPS vortex solutions were not discussed there, and they seem to be unaware of the fact that these are Alice strings even though in the same theory.}. 

The above action can be expanded in terms of component fields 
as
\begin{align}
\mathcal{L} &= -\frac{1}{4 e^2} f_{\mu\nu}^2  -\frac{1}{4g^2} F^a_{\mu\nu}F^a_{\mu\nu} + i{\lambda_0^\dagger} \bar\sigma^\mu\p_\mu \lambda_0 
+ i{\lambda^\dagger}^a \bar\sigma^\mu\left(\D_\mu \lambda\right)^a 
\nonumber\\
& + \sum_{i = \pm}\Tr\left[ \left(\D_\mu\Phi_i\right)^\dagger_a \D_\mu\Phi^i_a 
+\, {\bar\psi}^a_i \bar\sigma^\mu\left(\D\psi_i\right)^a + i \sqrt2 \Phi_i^\dagger \{\lambda,\psi_i\} -i \sqrt2  \{\bar\psi_i, \bar\lambda\}\Phi_i 
\right.\nonumber\\
&\left. \hspace{1cm}
+ D[\Phi_i, \Phi_i^\dagger] + F^\dagger_i F_i\right] 
\nonumber\\
& + \half D_a^2+ \half D_0^2  
+D_0 \Tr\left[\Phi_+^\dagger\Phi_+ - \Phi_-^\dagger\Phi_- -\xi^2\right] 
+\Tr\left[ i \sqrt2 \Phi_+^\dagger \psi_+\lambda_0 -i \sqrt2  \bar\psi_+\bar\lambda_0\Phi_+ \right] \nonumber\\& -\Tr\left[ i \sqrt2 \Phi_-^\dagger \psi_-\lambda_0 -i \sqrt2  \bar\psi_-\bar\lambda_0\Phi_- \right] .
\end{align}
The auxiliary fields $D, D^0$ and $F^\pm$ can be solved as
\begin{eqnarray}
\label{auxiliary}
D_0 = 2\xi^2 - \Tr \left(\Phi_+^\dagger\Phi_+ - \Phi_-^\dagger\Phi_-\right), \quad
D = - \half \sum_{i = \pm} \left[\Phi_i, \Phi_i^\dagger\right], \quad F^\pm =0.
\end{eqnarray}

If we set $\Phi_-=0$ and $\Phi_+ =\Phi$,
we then can recover the action in Eq.~(\ref{action}) 
in bosonic fields,
for which 
the potential term is 
\begin{eqnarray}
V(\Phi) =   \frac{g^2}{4} \Tr[\Phi,\Phi^\dagger]^2 
+  \frac{e^2}{2}\left(\Tr \Phi\Phi^\dagger  - 2 \xi^2\right)^2.
\end{eqnarray}
$\Phi_-$ parametrizes a flat direction of the theory.
The effect of a flat direction on BPS vortices of the ANO type was discussed in 
Refs.~\cite{Achucarro:2001ii, Penin:1996si}.

Now we show that
the BPS equations can be obtained by imposing SUSY transformation of the fermions to be zero. The SUSY transformations of the fermions can be written as 
\begin{eqnarray}
\delta_\epsilon\psi_\pm &=& i \sqrt 2 \sigma^\mu \bar\epsilon\D_\mu \Phi_\pm + \sqrt 2 \epsilon F_\pm, \\
\delta_\epsilon \lambda^a &=& \sigma^{\mu\nu}\epsilon F^a_{\mu\nu} + i \epsilon D^a,\\
\delta_\epsilon  \lambda^0 &=& i\epsilon D_0 + \sigma^{\mu\nu}\epsilon f_{\mu\nu} .
\end{eqnarray}
Here we have imposed the condition
\begin{eqnarray}\label{p0}
\langle \Psi_- \rangle = 0. 
\end{eqnarray}
 By  using above condition (\ref{p0}) and Eq.~(\ref{auxiliary}), 
we can recover the BPS equations by setting the SUSY transformation of the fermions to be zero:
\begin{eqnarray}
\delta_\epsilon\psi_+ &&= i \sqrt 2 \sigma^\mu \bar\epsilon\D_\mu \Phi_+ + \sqrt 2 \epsilon F_+ = 0 \nonumber \\ 
&&\Rightarrow \Big\{\left(D_1 + i D_2\right)\Phi_+\Big\}\bar\epsilon^{\dot 1} = \Big\{\left(D_1 - i D_2\right)\Phi_+\Big\}\bar\epsilon^{\dot 2} = 0 
\qquad\qquad
\end{eqnarray}
\begin{eqnarray}
\delta_\epsilon \lambda^a &&= \sigma^{\mu\nu}\epsilon F^a_{\mu\nu} + i \epsilon D^a = 0 \nonumber\\ &&\Rightarrow \left(F_{12} + \half \left[\Phi_+, \Phi_+^\dagger\right] \right)\epsilon_1= \left(F_{12} - \half \left[\Phi_+, \Phi_+^\dagger\right] \right)\epsilon_2 = 0 , 
\end{eqnarray}
and
\begin{eqnarray}
\delta_\epsilon  \lambda^0 &&= i\epsilon D_0 + \sigma^{\mu\nu}\epsilon f_{\mu\nu} = 0\nonumber \\ &&\Rightarrow \Big\{  f_{12} + e \left(\Tr \Phi_+\Phi_+^\dagger - 2 \xi^2\right)\Big\}\epsilon_1 = \Big\{  f_{12} - e \left(\Tr \Phi_+\Phi_+^\dagger - 2 \xi^2\right)\Big\}\epsilon_2 =0.
\end{eqnarray}
We then find that the condition
\begin{eqnarray}
\epsilon_2 = \epsilon^{\dot 2} = 0 
\end{eqnarray}
gives the BPS equations 
(\ref{eq:BPS1})--(\ref{eq:BPS3})
with the upper sign, while 
the condition
\begin{eqnarray}
\epsilon_1 = \epsilon^{\dot 1} = 0 
\end{eqnarray}
gives the same equations with the lower sign.
We thus have shown that a BPS Alice string 
preserves a half of SUSY and is a 1/2 BPS state.

\section{Summary and Discussion}
In this paper, we have presented a BPS construction of an Alice string in 
a non-Abelian gauge theory. To this end, we have considered a simple model 
of $U(1)\times SO(3)$ gauge theory with charged complex  scalar fields 
in the vector representation. 
To find a vortex solution, we first have written down the BPS equation 
by the Bogomol'nyi completion of the energy functional. 
We then have written our ansatz for the scalar and gauge fields 
for an axially symmetric single vortex. There are total four profile functions but BPS equations reduce them into two ($f_1(\rho), f_2(\rho)$). We have solved the profile functions numerically to construct solutions. 
We have found that equations for profile functions 
are the same with those of an axially symmetric non-Abelian vortex
in $U(N)$ gauge theory coupled with $N$ fundamental representation.
We then embed our Alice string in an ${\cal N}=1$ SUSY action by introducing two chiral superfields along with vector superfields corresponding to $U(1)$ and $SU(2)$ gauge theories with the Fayet-Iliopoulos term. 
The vortex is found to be 1/2-BPS saturated which follows directly from the ${\cal N}=1$ SUSY transformations.
In our Alice string construction the complex  adjoint scalar breaks the symmetry to $O(2)$. This $O(2)$ consists of a rotation around two different axis, that is, 
an $SO(2)$ rotation around $\tau^1$ and a $\pi$ rotation around $\tau^2$, 
$\tau^3$ or their linear combination ($\mathbb{Z}_2$). These two rotations actually do not commute with each other. This generates the ambiguity  in the presence of a vortex when the generator of $SO(2)$  becomes space dependent; once it encircles the Alice string, 
it rotates around $\tau^{2,3}$ by $\pi$ in the internal space. This implies that the element of $\mathbb{Z}_2$ acts non-trivially on the generator of the unbroken $SO(2)$, 
which changes the sign.

Alice strings found in this paper 
are the first explicit construction of BPS Alice strings in SUSY 
gauge theory.
It will be interesting to investigate what impact 
it has on various properties such as Cheshire charges 
in SUSY gauge theories.
It will be also possible to realize our SUSY gauge theory 
in D-brane configurations with an orientifold in string theory. It may be related to a D-brane configuration discussed in Ref~\cite{Harvey:2007ab, Harvey:2008zz}. It would be interesting to ask what are the consequences on D-brane physics of our BPS construction.\footnote{ Alice strings are  also discussed  in the context of defect branes in string theory \cite{Okada:2014wma}. 
}

Since the presence of our BPS Alice string breaks the $O(2)$ symmetry 
around the vortex core, it  produces a Nambu-Goldstone mode 
localized around the core giving rise to a $U(1)$ modulus. 
We have constructed the low energy effective theory explicitly in ref~\cite{Chatterjee:2017hya}.
In the SUSY context, there will be also fermion zero modes 
on the Alice string that constitute a SUSY multiplet of unbroken SUSY.
However, these modes will be 
non-normalizable as in Refs.~\cite{Alford:1990mk,Alford:1990ur},
while a relative $U(1)$ modulus of two Alice strings 
might be normalizable as in semi-local strings \cite{Eto:2007yv}
since two strings would not exhibit Alice properties.

In this paper, we have constructed only a single vortex solution.
The next step will be constructing multi-vortex configurations.
Arbitrary number of Alice strings should be able to be placed 
at any positions since our system is BPS and there should be no force among them.
We expect that odd number of Alice strings have Alice property 
while even number of them have no such property.
 
Our model admits a monopole since 
$\pi_2 [(S^1 \times S^2)/{\mathbb Z}_2] = {\mathbb Z}$.
In fact, a global monopole was constructed in spin-1 BEC
\cite{Ruostekoski:2003qx}.
The interesting is that this falls into 
an Alice string ring with twisted $U(1)$ modulus, 
as is the case of a monopole in the conventional Alice theory 
\cite{Benson:2004ue,Bais:2002ae}.
It is interesting question whether our model admits a BPS monopole  
and whether it is also a form of a twisted Alice ring.

In this paper we have considered only the simplest model with 
$U(1) \times SO(3)$ gauge theory with charged complex scalar fields 
in the vector representation or equivalently 
 $U(1) \times SU(2)$ gauge theory with charged complex adjoint scalar fields. 
 The simplest generalizations of our model will be charged complex scalar fields 
in any spin representation of $SO(3)$, 
$U(1) \times SO(N)$ gauge theory with charged complex scalar fields 
in the vector (or other) representation, 
and $U(1) \times SU(N)$ gauge theory with charged complex scalar fields 
in the adjoint representation.
In these cases, an unbroken non-Abelian gauge group in the bulk 
would have an obstruction 
in its continuos component 
(in addition to the Alice property in ${\mathbb Z}_2$), and 
it is broken in the vortex core giving rise to 
non-normalizable non-Abelian moduli as in Ref.~\cite{Bolognesi:2015ida}.

\section*{Acknowledgement}

This work is supported by the Ministry of Education, Culture, Sports, Science (MEXT)-Supported Program for the Strategic Research Foundation at Private Universities ``Topological Science'' (Grant No. S1511006). 
C.~C. acknowledges support as an International Research Fellow of the Japan Society for the Promotion of Science (JSPS)(Grant No: 16F16322). 
The work of M.~N. is supported in part by 
JSPS Grant-in-Aid for Scientific Research (KAKENHI Grant No. 25400268), 
and by a Grant-in-Aid for
Scientific Research on Innovative Areas ``Topological Materials
Science'' (KAKENHI Grant No.~15H05855) and ``Nuclear Matter in Neutron
Stars Investigated by Experiments and Astronomical Observations''
(KAKENHI Grant No.~15H00841) from the  Ministry of Education,
Culture, Sports, Science (MEXT) of Japan.

\end{document}